\documentclass[aps,pre,showpacs,groupedaddress,floatfix,nofootinbib]{revtex4}

\usepackage{amsmath}
\usepackage{graphicx}
\usepackage{color}

\begin{document}

\title{Comment on: ``Analytical approximations for the collapse of an empty
spherical bubble''}

\author{Paolo Amore}\email{paolo.amore@gmail.com}

\affiliation{Facultad de Ciencias, CUICBAS, Universidad de Colima, \\
Bernal D\'{\i}az del Castillo 340, Colima, Colima, Mexico}

\author{Francisco M. Fern\'andez}\email{fernande@quimica.unlp.edu.ar}

\affiliation{INIFTA (UNLP, CCT La Plata--CONICET), Blvd. 113 y 64 S/N, \\
Sucursal 4, Casilla de Correo 16, 1900 La Plata, Argentina}

\begin{abstract}
We analyze the Rayleigh equation for the collapse of an empty bubble and
provide an explanation for some recent analytical approximations to the
model. We derive the form of the singularity at the second boundary point
and discuss the convergence of the approximants. We also give a rigorous
proof of the asymptotic behavior of the coefficients of the power series
that are the basis for the approximate expressions.
\end{abstract}

\pacs{46.55.dp}

\maketitle

In a recent paper Obreschkow et al\cite{OBF12} derived simple and accurate
analytical approximations to the solution of the Rayleigh equation\cite{LR17}
for the collapse of an empty spherical bubble. Their approximants are based
on the expansion of the solution about the origin of time and can be
improved systematically. They showed that those simple analytical
expressions are suitable for the analysis of experimental cavitation data
obtained in microgravity.

Each approximant is the partial sum of the power series times a function
that takes into account the algebraic singularity at the other boundary
point. The authors also derived an approximate limit of the sequence of
partial sums in terms of the polylogarithm or Jonqui\`{e}re's function. To
this end, they resorted to a linear fit of the logarithm of the expansion
coefficients.

The results obtained by Obreschkow et al\cite{OBF12} are partly analytical
and partly numerical. In this comment we analyze them in a somewhat more
rigorous way with the purpose of providing a sound analytical foundation and
explanation of the main expressions.

It is sufficient for our purposes to restrict ourselves to the dimensionless
Rayleigh equation of motion for a collapsing bubble\cite{OBF12}
\begin{eqnarray}
r(t)\ddot{r}(t)+\frac{3}{2}\dot{r}(t)^{2}+\xi ^{2} &=&0,  \nonumber \\
r(0)=1,\,\dot{r}(0) &=&0  \label{eq:Rayleigh_dimensionless}
\end{eqnarray}
Taking into account the initial conditions and the fact that $r(-t)$ is also
a solution we conclude that $r(-t)=r(t)$. The equation (\ref
{eq:Rayleigh_dimensionless}) has been written in such a way that the bubble
collapses at $t=1$; that is to say: $r(1)=0$\cite{OBF12}.

If we multiply equation (\ref{eq:Rayleigh_dimensionless}) by $r^{2}\dot{r}$
and integrate with respect to the dimensionless time $t$ we obtain
\begin{equation}
\frac{3}{2}r^{3}\dot{r}^{2}+\xi ^{2}\left( r^{3}-1\right) =0
\label{eq:First_int}
\end{equation}
Note that $r(t)\equiv 1$ is a solution to Eq.~(\ref{eq:First_int}) that
satisfies the boundary conditions at $t=0$. Since this solution does not
satisfy Eq.~(\ref{eq:Rayleigh_dimensionless}) then both equations are not
identical. If we solve Eq.~(\ref{eq:First_int}) for $dt/dr$ and integrate
between $r=0$ and $r=1$ we obtain the value of $\xi $:\cite{LR17}
\begin{equation}
\xi =\sqrt{\frac{3}{2}}\int_{0}^{1}\frac{r^{3/2}\,dr}{\sqrt{1-r^{3}}}=\sqrt{%
\frac{3\pi }{2}}\frac{\Gamma \left( \frac{5}{6}\right) }{\Gamma \left( \frac{%
1}{3}\right) }\approx 0.914681  \label{eq:xi}
\end{equation}

If, on the other hand, we solve equation (\ref{eq:First_int}) for $\dot{r}%
^{2}$ and differentiate the result with respect to $t$ we obtain another
useful equation\cite{OBF12}
\begin{equation}
\ddot{r}=-\frac{\xi ^{2}}{r^{4}}  \label{eq:Ray_2}
\end{equation}

The solution to the Rayleigh equation (\ref{eq:Rayleigh_dimensionless}) can
be expanded in a Taylor series about the origin as follows:
\begin{eqnarray}
r(t) &=&\sum_{j=0}^{\infty }c_{j}t^{2j}  \nonumber \\
&=&1-\frac{\xi ^{2}t^{2}}{2}-\frac{\xi ^{4}t^{4}}{6}-\frac{19\xi ^{6}t^{6}}{%
180}-\frac{59\xi ^{8}t^{8}}{720}-\frac{4571\xi ^{10}t^{10}}{64800}-\cdots
\label{eq:t_series}
\end{eqnarray}
Although equations (\ref{eq:Rayleigh_dimensionless}), (\ref{eq:First_int})
and (\ref{eq:Ray_2}) are not identical this series can be obtained from any
of them and it converges for all $0\leq t\leq 1$ as discussed below.

Taking into account the initial conditions, the behavior of $r(t)$ about $%
t=1 $ and the symmetry of the solution Obreschkow et al\cite{OBF12} obtained
the first and simplest approximant $r_{0}(t)=\left( 1-t^{2}\right) ^{2/5}$.
This expression is considerably accurate in a neighborhood of $t=0$ because $%
\ddot{r}_{0}(0)=-0.4$ is quite close to $c_{1}\approx -0.418$. In addition
to it, the authors found that the error of this expression is smaller than $%
1\%$ for all $t$. For this reason they proposed the modified power-series
approximants
\begin{eqnarray}
r_{n}(t) &=&\left( 1-t^{2}\right) ^{2/5}S_{n}(t)  \nonumber \\
S_{n}(t) &=&\sum_{j=0}^{n}a_{j}t^{2j}=1+\frac{4-5\xi ^{2}}{10}t^{2}+\frac{%
42-30\xi ^{2}-25\xi ^{4}}{150}t^{4}+\ldots  \label{eq:r_n(t)}
\end{eqnarray}
where the functions $S_{n}(t)$ are the partial sums for the Taylor expansion
of $S(t)=r(t)/r_{0}(t)$ about $t=0$. Numerical calculation suggests that $%
a_{j}<0$ and $|a_{j+1}|<|a_{j}|$ for all $j$. Based on these results
Obreschkow et al\cite{OBF12} concluded that the approximants $r_{n}(t)$
converge monotonically towards $r_{\infty }(t)$ as $n\rightarrow \infty $.
The accuracy of these approximants increases with $n$, but according to
Obreschkow et al\cite{OBF12} $r_{\infty }(t)$ is not identical to $r(t)$
because $\ddot{r}(t)$ in Eq. (\ref{eq:Ray_2}) and $\ddot{r}_{\infty }(t)$
derived from Eq.~(\ref{eq:r_n(t)}) do not obey the same asymptotic behavior
as $t\rightarrow 1$.

Obreschkov et al\cite{OBF12} realized that $\ln (a_{j})$ vs $\ln (j)$ is an
almost straight line from which they estimated that $a_{j}\approx
a_{1}j^{-2.21}$. Based on this approximate relationship they derived the
following quite accurate analytical approximation to $r(t)$:
\begin{equation}
r_{*}(t)=\left( 1-t^{2}\right) ^{2/5}\left[ 1+a_{1}Li_{2.21}(t^{2})\right]
\label{eq:r*(t)}
\end{equation}
where $L_{s}(z)=\sum_{j=1}^{\infty }z^{j}/j^{s}$ is the polylogarithm or
Jonqui\`{e}re's function.

In what follows we will discuss the following points: first, why $r_{0}(t)$
and the approximants of greater order $r_{n}(t)$ are so accurate, second, if
$r_{\infty }(t)$ is equivalent to $r(t)$ for all $t$, and third, why $%
r_{\infty }(t)$ is approximately given by Eq.~(\ref{eq:r*(t)}). In order to
answer these questions we need the actual behavior of $r(t)$ as $%
t\rightarrow 1$.

We can obtain the asymptotic behavior of $r(t)$ as $t\rightarrow 1$ most
easily from Eq.~(\ref{eq:First_int}); the result is
\begin{equation}
r(t)=\frac{\left[ 180\xi \left( 1-t\right) \right] ^{2/5}}{6}-\frac{5\times
750^{1/5}\left[ \xi \left( 1-t\right) \right] ^{8/5}}{66}+O((1-t)^{14/5})
\label{eq:r_asympt}
\end{equation}
It is worth noting that the leading term
\begin{equation}
r_{a}(t)\equiv \frac{\left[ 180\xi (1-t)\right] ^{2/5}}{6}\approx
1.28371(1-t)^{2/5}  \label{eq:r_a(t)}
\end{equation}
is an exact solution to Eq.~(\ref{eq:Ray_2}) that does not satisfy the
initial conditions. The function (\ref{eq:r_a(t)}) does not satisfy the
other two alternative equations (\ref{eq:Rayleigh_dimensionless}) and (\ref
{eq:First_int}).

If we substitute $r(t)=r_{0}(t)S(t)$ into either of the equations (\ref
{eq:Rayleigh_dimensionless}), (\ref{eq:First_int}) or (\ref{eq:Ray_2}) and
take the limit $t\rightarrow 1^{-}$ then we obtain
\begin{equation}
S(1)=\frac{\left( 90\xi \right) ^{2/5}}{6}\approx 0.972867  \label{eq:S(1)}
\end{equation}
that is consistent with the analytical asymptotic expression (\ref
{eq:r_asympt}) as follows from
\begin{equation}
\lim_{t\rightarrow 1^{-}}\frac{r(t)}{r_{0}(t)}=\lim_{t\rightarrow 1^{-}}%
\frac{r_{a}(t)}{r_{0}(t)}=\frac{\left( 90\xi \right) ^{2/5}}{6}
\label{eq:r(t)/r0(t)}
\end{equation}

Obreschkow et al\cite{OBF12} already proved that $r_{0}(t)$ is a reasonably
good approximation to $r(t)$ in the neighborhood of $t=0$. Eq.~(\ref
{eq:r(t)/r0(t)}) tells us that $r_{0}(t)$ is also quite close to $r(t)$ in
the neighborhood of $t=1$. For this reason $r_{0}(t)$ is so accurate for all
$t$ and the approach of Obreschkow et al\cite{OBF12} is remarkably
successful even when the sequence of partial sums $S_{n}(t)$ converges
slowly.

Let us now go into the question whether $r_{n}(t)$ actually gives $r(t)$
when $n\rightarrow \infty $. To begin with, note that the sequence of
partial sums converges for all $t<1$ because the singular point closest to
the origin is located at $t=1$. Therefore it is clear that if $S_{n}(1)$
converges towards $S(1)$ as $n\rightarrow \infty $ then $S_{n}(t)$ converges
towards $S(t)$ for all $t$ and $r_{n}(t)$ converges towards the actual
solution $r(t)$ of the dimensionless Rayleigh equation. Our numerical
analysis suggests that $S_{n}(1)\rightarrow $ $S(1)$ as $n\rightarrow \infty
$; compare, for example, $S_{200}(1)\approx 0.972892$ with Eq.~(\ref{eq:S(1)}%
). If we accept that $S_{\infty }(1)=S(1)$ then we can easily prove that $%
r_{\infty }(t)$ satisfies any of the equations (\ref
{eq:Rayleigh_dimensionless}), (\ref{eq:First_int}) or (\ref{eq:Ray_2}) as $%
t\rightarrow 1$. Consider, for example, Eq.~(\ref{eq:Ray_2}). If we
substitute $r(t)=r_{0}(t)S(t)$ then $\lim_{t\rightarrow 1^{-}}r(t)^{4}\ddot{r%
}(t)=-24S(1)^{5}/25=-\xi ^{2}$. On the other hand $\lim_{t\rightarrow
1^{-}}r_{n}(t)^{4}\ddot{r}_{n}(t)=-24S_{n}(1)^{5}/25$ which proves the
point. Therefore, if $r_{\infty }(t)$ satisfies Eq.~(\ref{eq:Ray_2}) for the
most unfavorable case $t=1$ then it satisfies that equation for all $t$.

The approximant (\ref{eq:r*(t)}) is quite accurate in the neighborhood of $%
t=1$ because
\begin{equation}
\lim_{t\rightarrow 1}\frac{r_{*}(t)}{(1-t)^{2/5}}=2^{2/5}\left[
1+a_{1}Li_{2.21}(1)\right] \approx 1.28363
\end{equation}
is very close to the exact asymptotic behavior given by Eq.~(\ref{eq:r_a(t)}%
). In what follows we show how the form of $r_{*}(t)$ emerges from the
asymptotic behavior of the coefficients $a_{j}$.

To begin with, note that if $f(x)$ exhibits a branch point at $x=x_{0}$ with
exponent $\alpha $ $(1-x/x_{0})^{\alpha }$ then the coefficients $c_{j}$ of
the Taylor expansion about $x=0$ for $f(x)$ behave asymptotically as $%
|c_{j}|\sim c|x_{0}|^{-j}j^{-\alpha -1}$, where $c$ is a constant.
Obviously, we are assuming that there is no other singularity closer to the
origin or in the vicinity of $x_{0}$. For this reason, the coefficients of
the original series (\ref{eq:t_series}) behave approximately as $%
c_{j}\approx cj^{-7/5}$ reflecting the branch-point singularity at $t=1$
with exponent $\alpha =2/5$. Note that the Taylor series about $x=0$ for $%
f(x)$ converges for all $0\leq x\leq x_{0}$ if $\alpha >0$ and that the rate
of convergence increases with $\alpha $. The function $S(t)$ exhibits a
branch-point singularity at $t=1$ with exponent $\alpha =6/5$ as shown by
the asymptotic expansion
\begin{equation}
S(t)=\frac{r(t)}{r_{0}(t)}=\frac{90^{2/5}\xi ^{2/5}}{6}+\frac{90^{2/5}\xi
^{2/5}}{30}(1-t)-\frac{5\times 6000^{1/5}\xi ^{8/5}}{132}%
(1-t)^{6/5}+O((1-t)^{2})  \label{eq:S(t)_asympt}
\end{equation}
Therefore, the coefficients $a_{j}$ behave asymptotically as $|a_{j}|\sim
aj^{-6/5-1}=aj^{-2.2}$, where $a$ is a constant. This theoretical result
clearly explains the outcome of the linear fitting by which Obreschkow et al%
\cite{OBF12} obtained the approximant $r_{*}(t)$. The slight discrepancy
between the theoretical and numerical exponents is due to the fact that
those authors fitted all the coefficients $a_{j}$ and the asymptotic
behavior is determined by those of sufficiently large $j$. If, for example,
we fit the coefficients $a_{j}$ for $100\leq j\leq 150$ then we obtain a
much better agreement between theory and numerical approximation: $%
|a_{j}|\approx 0.017j^{-2.20}$. Obviously, the reason for fitting all the
coefficients is the practical purpose of obtaining a suitable approximation
for all $t$\cite{OBF12}. In the present case we are mainly interested in
explaining the form of the approximant (\ref{eq:r*(t)}) and for that reason
we resort to fitting the coefficients with the largest available orders that
reflect the asymptotic behavior of $r(t)$ close to $t=1$. We also appreciate
that the sequence of partial sums $S_{n}(t)$ converges for all $0\leq t\leq
1 $ because $1+\alpha =11/5>1$ and that the rate of convergence of the
series (\ref{eq:r_n(t)}) is greater than the one of (\ref{eq:t_series}).

Finally, we want to discuss an alternative power series with much better
convergence properties. It is well known that in some cases the inverted
series exhibits better convergence properties than the original one\cite
{AF08}. The series inversion is the basis for the parametric perturbation
theory\cite{A07}. In the present case we define the new variable $\rho $
\begin{eqnarray}
\rho &=&\frac{r-1}{c_{1}}=\frac{2}{\xi ^{2}}(1-r)  \nonumber \\
&=&z+\frac{\xi ^{2}z^{2}}{3}+\frac{19\xi ^{4}z^{3}}{90}+\frac{59\xi ^{6}z^{4}%
}{360}+\frac{4571\xi ^{8}z^{5}}{32400}+\ldots  \label{eq:rho}
\end{eqnarray}
where $z=t^{2}$, and invert the series to obtain $z(\rho )$:
\begin{eqnarray}
z &=&\rho +\sum_{j=2}^{\infty }b_{j}\rho ^{j}  \nonumber \\
&=&\rho -\frac{\rho ^{2}\xi ^{2}}{3}+\frac{\rho ^{3}\xi ^{4}}{90}+\frac{\rho
^{4}\xi ^{6}}{360}+\frac{7\rho ^{5}\xi ^{8}}{10800}+\ldots
\label{eq:rho_series}
\end{eqnarray}
It follows from the asymptotic expansion

\begin{equation}
t=1-\frac{\sqrt{6}r^{5/2}}{5\xi }-\frac{\sqrt{6}r^{11/2}}{22\xi }+\ldots
\label{eq:t(r)_asympt}
\end{equation}
that $z(\rho )$ exhibits a singularity of the form $\left( 1-\rho /\rho
_{0}\right) ^{5/2}$, where $\rho _{0}=$ $2/\xi ^{2}\approx 2.3905$.
Therefore, the coefficients $b_{j}$ behave asymptotically as $|b_{j}|\sim
b|\rho _{0}|^{-j}j^{-7/2}$, where $b$ is a positive constant. It follows
from fitting $\ln (|b_{j}|)$ for $80\leq j\leq 100$ that $|b_{j}|\sim
1.78\times 2.39^{-j}\times j^{-3.6}$, where $3.6\approx $ $7/2$ and $%
2.39\approx 2/\xi ^{2}$ which confirm the theoretical result.

Clearly, the coefficients of the inverted series decrease more
rapidly than the coefficients of either (\ref{eq:t_series}) or
(\ref{eq:r_n(t)}). Therefore, from a numerical point of view it is
convenient to build approximants based on the inverted series. The
price we have to pay is that the inverted series does not yield
$r(t)$ directly, which may not be a serious drawback for some
purposes. We can improve the convergence of the inverted series by
means of an appropriate summation method like the Pad\'{e}
approximants and thus obtain $t(\rho )=\sqrt{z(\rho )}$ which
together with $r(\rho )=1-\frac{\xi ^{2}\rho }{2}$ yields the
parametric representation for $r(t)$.

Summarizing: the most important results of this comment are

\begin{itemize}
\item  The particular form and remarkable accuracy of the approximation
$r_{*}(t)$ of Obreschkow et al is now explained by the fact that the
coefficients $a_{j}$ behave asymptotically as $j^{-11/5}$ for large $j$.

\item  Present analysis strongly suggests that $r_{\infty }(t)$ is identical to
$r(t)$ for all $0\leq t\leq 1$ in disagreement with the statement
of breschkow et al.

\item  Alternative approximations to $r(t)$ in terms of the inverted power
series exhibit faster convergence than the approach of Obreschkow et al,
although the inverted power series does not yield $r(t)$ directly.
\end{itemize}

\acknowledgements

P. Amore acknowledges support of Conacyt through the SNI fellowship and also
of PIFI. F.M.F acknowledges support of PIFI and of UNLP through the
``subsidio para viajes y/o estad\'{i}as''. We also thank Professor
Obreschkow for valuable comments.


\begin{thebibliography}{9}
\bibitem{OBF12}  D. Obreschkow, M. Bruderer, and M. Farhat, Phys. Rev. E
\textbf{85}, 066303 (2012).

\bibitem{LR17}  Lord Rayleigh, Philos. Mag. \textbf{34}, 94 (1917).

\bibitem{AF08}  P. Amore and F. M. Fern\'{a}ndez, J. Phys. A \textbf{41},
025201 (2008).

\bibitem{A07}  P. Amore, Phys. Rev. D \textbf{76}, 076001 (2007).
\end{thebibliography}
\end{document}